\input mtexsis
\texsis
\paper

\def\ct{\hbox{T}}

\def\hct{\hat{\ct}}

\titlepage
\hbox{\space}\vskip 1in
\title Particle Motion in the Stable Region Near an Edge of a 
Linear Half-Integer Stopband
\endtitle
\author
George Parzen
July 20, 1995
BNL-62036
\endauthor
\abstract
	This paper studies the motion of a particle whose tune is inside
and near a linear half-integer stopband.  Results are found for the tune
and beta functions in the stable region close to an edge of the stopband.  
It is shown that the eigenvalues and
the eigenfunctions of the transfer matrix are real inside the
stopband.  All the results found are also valid for small accelerators 
where the large accelerator approximation is not used.
\endabstract
\endtitlepage

\section{Introduction}

Inside a linear-half integer stopband the particle motion can be
unstable and grow exponentially.  A result is found for the
growth parameter $g$ which is the imaginary part of the tune.  It
is shown that the real part of the tune inside the stopband is
constant at the value $q/2$, where $\nu=q/2$ is the center of
the stopband, $q$ being some integer.  The proof given does not
depend on perturbation theory, and the result follows from the
symplectic properties.  The eigenvalues and eigenfunctions of the
transfer matrix are shown to be real inside the stopband.  In the
stable region near an edge of the stopband, the tune varies
rapidly and the beta function becomes infinite as the unperturbed
tune approaches the edge of the stopband.  Results are found
for the tune and beta function in the stable region near an edge
of the stopband.  It is found that the beta function becomes
infinite inversely as the square root of the distance of the
unperturbed tune from the edge of the stopband.  The basic equations
used are also valid for small accelerators, where the large
accelerator approximations is not used, and all the results found
are valid for small accelerators.

\section{Results when the tune is not in the stopband}

Before treating the interesting case where the tune is inside the
stopband, it will be helpful to first treat the case where the
tune is not inside the stopband.  It will then become clear
where the perturbation solution breaks down, when the tune is
inside the stopband, and how the perturbation solution can be
repaired.

It will be assumed that in the absence of the perturbing fields,
the tune of the particle is $\nu_0$ and that the motion is stable
when $\nu_0$ is close to $q/2$, where $q$ is an integer.

It is assumed that a perturbing field is present which is given
on the median plane by
$$\Delta B_y = - G(s) x \eqno{\hbox{(2-1)}}$$
$G(s)$ is periodic in $s$ and contains the field harmonics
that can excite the stopband around $\nu_0=q/2$.

Introducing $\eta$ defined by
$$\eta = x/\beta^{1/2} , \eqno{\hbox{(2-2)}}$$
where $\beta$ is the beta function of the unperturbed field, the
equation of motion can be written as
$$\eqalign{ 
{d^2\eta\over d\theta^2} &+ \nu_0^2 \eta = f \cr
f &= \nu_0^2 \beta^{3/2} \Delta B_y/B\rho \cr
f &= -\nu_0^2 \beta^2 G \eta/B\rho \cr
B\rho &= pc/e , \ d\theta = ds/\nu_0\beta \cr} \eqno{\hbox{(2-3)}}$$
Eqs. (2-3) are valid for large accelerators, and a small change is
required$^1$ to make them valid for small accelerators (see section 7).
However, the final results found below are valid for small accelerators
that require the use of the exact linearized equations.  This is shown
in section 7.

Eq. (2-3) can be written as
$$\eqalign{
{d^2\eta\over d\theta^2} + \nu_0^2 &= -2\nu_0 b(\theta) \eta \cr
b(\theta) &= {1\over 2} \nu_0 \beta^2 G/B\rho \cr} \eqno{\hbox{(2-4)}}$$
Because $b(\theta)$ is periodic a solution for $\eta$ will have the
form
$$\eta = \exp (i\nu_s \theta) h(\theta) \eqno{\hbox{(2-5)}}$$
where $h(\theta)$ is periodic.  It is assumed that the tune $\nu_0$
will change to $\nu_s$ because of the perturbing field.  Thus $\eta$
can be assumed to have the form
$$\eqalign{
\eta &= A_s \exp (i\nu_s\theta) + \sum_{r\ne s} A_r \exp (i\nu_r\theta) \cr
\nu_r &= \nu_s + n \cr} \eqno{\hbox{(2-6)}}$$
where $n$ is some integer but $n\ne 0$.  For a zero perturbing field
the solution for $\eta$ is $\eta=A\exp(i\nu_0\theta)$.  Thus for
small perturbing fields it can be assumed that
$$\eqalign{
\nu_s &\simeq \nu_0 \cr
A_r &\ll A_s \ \hbox{for}\ r\ne s \cr} \eqno{\hbox{(2-7)}}$$
Putting Eq. (2-6) into Eq. (2-4), one obtains a set of equations
for the $A_r$
$$\eqalign{
(\nu_r^2-\nu_0^2) A_r &= 2\nu_0 \sum_{\overline{r}} b_{r\overline{r}}
A_{\overline{r}} \cr
b_{r\overline{r}} &= {1\over 2\pi} \int_0^{2\pi} d\theta \ b(\theta)
\exp(-i\nu_r\theta+i\nu_{\overline{r}}\theta) \cr
\nu_r &= \nu_s + n \cr} \eqno{\hbox{(2-8)}}$$

Eqs. (2-8) are a set of homogenious equations for the $A_r$.  The
condition required for a solution to exist is that the determinant of
the coefficients of the $A_r$ should vanish.  This condition will
determine $\nu_s$, and then Eq. (2-8) can be solved for the $A_r$
in terms of $A_s$.  It is more convenient to solve Eqs. (2-8) using
an iterative perturbation procedure.  For the initial guess for $\eta$
in this iterative procedure, one can assume
$$\eta = A_s \exp (i\nu_s\theta) \eqno{\hbox{(2-9)}}$$
One can put this result for $\eta$ in the right hand side of Eq.
(2-8), and solve for $A_r$ which gives 
$$\eqalign{
\left( \nu_r^2-\nu_0^2\right) A_r &= 2\nu_0 b_{rs} A_s \cr
\nu_r &= \nu_s + n \cr} \eqno{\hbox{(2-10)}}$$
For $r=s$, Eq. (2-10) gives
$$\left( \nu_s^2-\nu_0^2\right) A_s = 2\nu_0 b_{ss} A_s 
\eqno{\hbox{(2-11)}}$$
which determines $\nu_s$.  One finds
$$\nu_s - \nu_0 = b_{ss} \eqno{\hbox{(2-12a)}}$$
which can be written as
$$\eqalign{
\nu_s - \nu_0 &= b_0 \cr
b_0 &= {1\over 2\pi} \int_0^{2\pi} d\theta b(\theta) \cr
b_0 &= {1\over 4\pi} \int_0^L \beta G/B\rho \cr} \eqno{\hbox{(2.12b)}}$$
This is the well known result$^{2,3}$ for the first order tune shift due
to a gradient $G$.  The solution for $\eta$ to first order can be
found using Eq. (2-6) and Eq. (2-8)
$$\eqalign{
\eta &= A_s \left\{ \exp(i\nu_s\theta) + \sum_{r\ne s}{2\nu_0\over\nu_r^2 
- \nu_0^2} b_{rs} \exp(i\nu_r\theta)\right\} \cr
\nu_r &= \nu_s + n , \qquad n\ne 0 \cr} \eqno{\hbox{(2-13a)}}$$
which can be also written as
$$\eqalign{
\eta &= A_s \exp (i\nu_s\theta) \left\{ 1+ \sum_{n\ne 0} {2\nu_0 b_n \over
n(n+2\nu_0)} \exp (in\theta) \right\} \cr
b_n &= {1\over 4\pi} \int_0^L ds \beta G \exp(-in\theta)/B\rho \cr}
\eqno{\hbox{(2-13b)}}$$
One can see from Eq. (2-13b) that the above solution will not be
valid when $\nu_0$ is close to a half-integer for the denominator
$(n+2\nu_0)$ can become very small.

\section{Results when the tune is inside the stopband}

It is assumed that $\nu_0$ is near $\nu_0=q/2$, $q$ being an integer.
The stopband is defined as the range of $\nu_0$ for which the tune,
$\nu_s$, in the presence of the perturbation given by Eq. (2-2) has a
non-zero imaginary part.  One can write $\nu_s$ as
$$\nu_s = \nu_{sR} - ig \eqno{\hbox{(3-1)}}$$
It will be shown that inside the stopband, where $g\ne 0$, then
$$\nu_{sR} = q/2 \eqno{\hbox{(3-2)}}$$
This may be shown as follows.  Let $\mu$ be the phase shift for a
period, and $\mu=2\pi\nu_s$ where the period has been assumed to be
one turn or $2\pi$ in $\theta$.  Let $T$ be the transfer matrix for
one period.  Then one has
$$\cos\mu = {1\over 2} (T_{11} + T_{22}), \eqno{\hbox{(3-3)}}$$
and one sees that $\cos\mu$ is real even inside the stopband since
the $T_{ij}$ are real.  One also has
$$\cos\mu=\cos2\pi(\nu_{sR}-ig)=\cos(2\pi\nu_{sR})\cosh(2\pi g) +
i\sin(2\pi\nu_{sR})\sinh 2\pi g \eqno{\hbox{(3-4)}}$$
In order for $\cos \mu$ to be real, one has to have either
$g=0$ or, if $g\ne 0$, $2\pi\nu_{sR}=n\pi$, where $n$ is some
integer.  Thus, inside the stopband where $g\ne 0$,
$$\nu_{sR} = n/2 \eqno{\hbox{(3-5)}}$$
In order to have continuity when the perturbation goes to zero,
one has $n=q$ and
$$\nu_{sR} = q/2 \eqno{\hbox{(3-6)}}$$
For the range of $\nu_0$ which is inside the stopband, $\nu_{sR}$
remains constant at $\nu_{sR}=q/2$ while $g$ goes, as will be
seen below, from $g=0$ at the edge of the stopband to a maximum
value in the middle of the stopband, where $\nu_0=q/2$.  One may
note that this result does not depend on the use of perturbation
theory and is valid even for large perturbations.  A proof of the
result, based on perturbation theory, is given below.

Now let us return to the problem of computing the growth factor,
$g$, using perturbation theory.  The iterative perturbation approach
used in section 2 appears to breakdown when $\nu_0$ is close to
$q/2$.  One sees from Eq. (2-10) that when $\nu_0\simeq q/2$, then
one of the $A_r$ becomes comparable to $A_s$, and this is the $A_r$
for which $\nu_r=\nu_s-q$, $\nu_r\simeq -q/2$.  Thus in the above
iterative procedure for finding $\eta$, one will assume for the
initial guess for $\eta$, instead of Eq. (2-9),
$$\eqalign{
\eta &= A_s \exp (i\nu_s\theta) + A_{\overline{s}} \exp(i\nu_{\overline{s}}
\theta) \cr
\nu_{\overline{s}} &= \nu_s - q \cr} \eqno{\hbox{(3-7)}}$$
Then Eq. (2-10) is replaced by
$$\eqalign{
(\nu_r^2-\nu_o^2) A_r &= 2\nu_0 b_{rs} A_s + 2\nu_0 b_{r\overline{s}}
A_{\overline{s}} \cr
\nu_r &= \nu_s+n \quad \hbox{or} \quad \nu_r = \nu_{\overline{s}}+n =
\nu_s-q+n \cr} \eqno{\hbox{(3-8)}}$$
For $r=s$ and $r=\overline{s}$ one obtains 2 equations for $A_s$ and
$A_{\overline{s}}$
$$\eqalign{
(\nu_s^2-\nu_0^2) A_s &= 2\nu_0 b_{s\overline{s}} A_{\overline{s}} \cr
(\nu_{\overline{s}}^2-\nu_0^2) A_{\overline{s}} &= 2\nu_0 b_{\overline{s}s}
A_s \cr
\nu_{\overline{s}} &= \nu_s - q \cr} \eqno{\hbox{(3-9)}}$$
In Eq. (3-9), it has been assumed, for simplicities sake, that
$b_{ss}=0$.  This can be accomplished by redefining $\nu_0$ to be
$\nu_0+b_0$.

In order for Eqs. (3-9) to have a solution, one must have
$$\eqalign{
(\nu_s^2-\nu_0^2) (\nu_{\overline{s}}^2 - \nu_0^2) &= 4\nu_0^2
|b_{s\overline{s}}|^2 \cr
\nu_{\overline{s}} &= \nu_s - q \cr} \eqno{\hbox{(3-10)}}$$
Eq. (3-10) determines $\nu_s$.  If one writes $\nu_s=\nu_{sR} - ig$,
one finds 
$$\left[ (\nu_{sR} - ig)^2-\nu_0^2 \right] \left[ (\nu_{sR} - q - ig)^2
\right] = 4\nu_0^2 |\Delta\nu|^2 \eqno{\hbox{(3-11a)}}$$
where $\Delta\nu = b_{s\overline{s}}$
$$\Delta\nu = {1\over 4\pi} \int_0^L ds\ \beta \exp(-iq\theta)
G/B\rho  \eqno{\hbox{(3-11b)}}$$
$|\Delta\nu|$ will turn out to be the half width of the stopband.  
Assuming that $\nu_0$ and $\nu_{sR}$ are close to $q/2$, then Eq. (3-11a)
can be rewritten as
$$(\nu_{sR} - ig - \nu_0) (q-\nu_{sR} + ig - \nu_0) = |\Delta\nu|^2
\eqno{\hbox{(3-12)}}$$
The imaginary part of the left hand side gives the equation,
$$g(2\nu_{sR} - q) = 0 \eqno{\hbox{(3-13)}}$$
Inside the stopband where $g\ne 0$, one gets
$$\nu_{sR} = q/2 \eqno{\hbox{(3-14)}}$$
Using Eq. (3-14) for $\nu_{sR}$, the real part of Eq. (3-12) gives
$$\eqalign{
(q/2-\nu_0)^2 + g^2 &= |\Delta\nu|^2 \cr
g &= \pm \left\{|\Delta\nu|^2-(q/2-\nu_0)^2\right\}^{1/2} \cr}
\eqno{\hbox{(3-15)}}$$
Eq. (3-15) shows that the growth parameter $g$ has a maximum at
$\nu_0=q/2$ where $g=|\Delta\nu|$, and decreases to zero at
$\nu_0 = q/2\pm|\Delta\nu|$.  The stopband width is then
$2|\Delta\nu|$ and extends from $q/2-|\Delta\nu|$ to $q/2+|\Delta\nu|$.
Thus $|\Delta\nu|$ is the half-width of the stopband.

Now let us find the solutions for $\eta$ that will give the particle
motion inside the stopband.  To lowest order, $\eta$ is given by
Eq. (3-7) and inside the stopband
$$\eqalign{
\eta &= A_s \exp (i\nu_s\theta) + A_{\overline{s}} \exp (i\nu_{\overline{s}}
\theta) \cr
\nu_s &= q/2 -ig \cr
\nu_{\overline{s}} &= -q/2 -ig \cr} \eqno{\hbox{(3-16)}}$$
$A_s$ and $A_{\overline{s}}$ are related by Eqs. (3-9) which gives
$$\eqalign{
A_{\overline{s}} &= {2\nu_0 b_{\overline{s}s}\over(q/2+ig)^2-\nu_0^2}
A_s \cr
A_{\overline{s}} &= {b_{\overline{s}s}\over (q/2-\nu_0)+ig} A_s \cr}
\eqno{\hbox{(3-17)}}$$
where it has been assumed that $\nu_0$ is close to $q/2$.  Eq. (3-17) can
be written as
$$\eqalign{
A_{\overline{s}} &= A_s \exp[-i(\delta_1+\delta_2)] \cr
\delta_1 &= \hbox{phase} \ \Delta\nu \cr
\delta_2 &= \hbox{phase} \ [(q/2-\nu_0)+ig] \cr} \eqno{\hbox{(3-18)}}$$
where $b_{\overline{s}s}=|\Delta\nu|\exp (-i\delta_1)$, and 
$|\Delta\nu|=g^2+(q/2-\nu_0)^2$ were used.

Putting these results for $A_{\overline{s}}$ into Eq. (3-16) one gets
for $\eta$ and $x$
$$\eqalign{
\eta &= A_s \exp(g\theta) \cos(q\theta/2-(\delta_1+\delta_2)/2) \cr
\delta_1 &= \hbox{phase}\ \Delta\nu \cr
\delta_2 &= \hbox{phase}\ [(q/2-\nu_0)+ig] \cr
x &= \beta^{1/2} \eta \cr} \eqno{\hbox{(3-19)}}$$
where the multiplying constant $\exp[-i(\delta+\delta_2)/2]$ was
dropped.  Eq. (3-19) give to lowest order the two solutions inside
the stopband corresponding to whether $g$ is positive or negative.
The first order correction to $\eta$ could also be found using
Eq. (3-8) and the result is
$$\eqalign{
\eta &= A_s \exp(g\theta) \left[\cos(q\theta/2-(\delta_1+\delta_2)/2) 
\phantom{\sum_{n\ne0}} \right. \cr
&+ \left. \sum_{n\ne 0,-q} {2\nu_0\over n(n+2\nu_0)} b_n \cos((q/2+n)\theta -
(\delta_1+\delta_2)/2)\right] \cr
b_n &= {1\over 4\pi} \int_0^L ds\ \beta \ G \ \exp(-in\theta)/B\rho \cr}
\eqno{\hbox{(3-20)}}$$
The results found above for the stopband width, the growth rate and
particle motion in the stopbands will apply equally well to small
accelerators where the exact linear equations have to be used, without
the approximations used for large accelerators, as shown in section 7.

\section{The realness of the eigenvalues and eigenvectors inside the
stopband}

It will be shown in this section that the eigenvalues and the 
eigenvectors of the one period transfer matrix are real when $\nu_0$
is inside the stopband.  If $\hct(s)=\ct(s+L,s)$ is the one period
transfer matrix, then the eigenvalues and eigenvectors satisfy the
equation
$$\eqalign{
\hct(s) \ x(s) &= \lambda \ x(s) \cr
x &= \left[ \matrix{ x \cr p_x \cr} \right] \cr} \eqno{\hbox{(4-1)}}$$
The symbol $x$ is used to indicate both the column vector $x$ and
the first element of this vector.  The meaning of $x$ should be clear
from the context.  The eigenvalue $\lambda$ is related to $\nu$-value
by
$$\eqalign{
\lambda &= \exp(i\mu) \cr
\mu &= 2\pi\nu \cr} \eqno{\hbox{(4-2)}}$$
where, for simplicity, the period is assumed to be one turn or
$2\pi$ in $\theta$.

In section 3 it was shown that when $\nu_0$ is inside the stopband,
$\nu$ can be written as $\nu=q/2-ig$ one finds for $x$, when $\nu_0$
is inside the stopband,
$$\eqalign{
\lambda &= \exp (iq\pi+2\pi g) \cr
\lambda &= (-1)^q \exp(2\pi g) \cr} \eqno{\hbox{(4-3)}}$$
and the eigenvalue, $\lambda$, is real when $\nu_0$ is inside the 
stopband.

Now let us consider the eigenvectors of $\hct$.  If $\lambda$, $x$
are an eigenvalue and eigenvector of $\hct$, it is known that $1\lambda$
is an eigenvalue of $\hct$, as $\hct$ is symplectic.  Since $\hct$ is
real, $\lambda^*$, $x^*$ are also an eigenvalue and eigenvector
respectively of $\hct$.  It appears that one has 3 eigenvalues,
$\lambda, 1/\lambda$ and $\lambda^*$ instead of just two.  Two of
the 3 apparent eigenvalues must be equal.

When $\nu_0$ is not in the stopband then $\nu_s$ is real, 
$1/\lambda = \exp(-i\mu)=\lambda^*$, and the second solution is
$\lambda^*, x^*$.

When $\nu_0$ is in the stopband, then $\nu$ has an imaginary part,
$\nu=q/2-ig$.  Then $1/\lambda\ne\lambda^*$. To avoid having 3
eigenvalues, one needs to have $\lambda^*=\lambda$ and $x^*=x$, or
$\lambda$ and $x$ are real when $\nu_0$ is in the stopband.

One should note that the statement $x$ is real means here that $x$
can be made real by multiplying by a constant.  This is because an
eigenvector can be multiplied by a constant and still be an eigenvector
of $\hct$.  Thus the eigenvector may not appear real because of the
presence of a constant complex multiplier, but can be made real by
multiplying it with a constant.

In section 3, where $\eta$ was found by an iterative perturbation
procedure for $\nu_0$ in the stopband, the solution found, Eq. (3-17),
is real as required by the above theorm, whereas when $\nu_0$ is not
inside the stopband, the solution for $\eta$ found Eq. (2-13) is
complex.  Since the solution for $\eta$ was specified to have the
form given by Eq. (2-6), then $x=\eta/\beta^{1/2}$ is the first
element of the eigenvector, $x$, of $\hct$.

\section{Tune near the edge of a stopband}

In this section, a result will be found for the tune in the stable
region outside the stopband but close to one of the edges of the
stopband.  It will be shown that close to the edge of a stopband,
$$|\nu-q/2| = \left\{ 2|\Delta\nu|\ |\nu_0-\nu_e| \right\}^{1/2}
\eqno{\hbox{(5-1)}}$$
$\nu$ is the tune in the presence of the gradient perturbation,
$\nu_e$ is the edge of the stopband, $\nu_e=q/2\pm|\Delta\nu|$.
$|\Delta\nu|$ is the half-width of the stopband.  Eq. (5-1) shows 
that when $\nu_0$ is close to an edge of the stopband, $\nu_e$,
$\nu$ varies rapidly with $\nu_0$, and the slope of the $\nu$ vs.
$\nu_0$ curve is vertical at $\nu_0=\nu_e$.

To find $\nu$ in the stable region outside the stopband, where
$\nu_0-q/2|>|\Delta\nu|$, one goes back to the derivation given
in section 3 for $\nu$ inside the stopband, starting with Eq. (3-12).
Eq. (3-15) shows that for $|\nu_0-q/2|>|\Delta\nu|$ the only
acceptable solution is $g=0$, and Eq. (3-12) can be written as
$$(\nu-\nu_0) (|\nu-q|-\nu_0) = |\Delta\nu|^2 \eqno{\hbox{(5-2)}}$$
where we have put $\nu_s=\nu$.

Assuming that $\nu_0$ is just below the stopband edge $\nu_e=q/2-
|\Delta\nu|$, put $\nu_0=\nu_e-\epsilon$ and $\nu=q/2-\delta$ into
Eq. (5-2), where $\epsilon$ and $\delta$ both approach zero as
$\nu_0$ approaches the stopband edge.  We find
$$\delta = \left\{ \epsilon(\epsilon+2|\Delta\nu|)\right\}^{1/2}
\eqno{\hbox{(5-3)}}$$
The top edge of the stopband can be treated in the same way and both
results can be combined into the one result
$$\eqalign{
|\nu-q/2| &= \left\{|\nu_0-\nu_e| (|\nu_0-\nu_e|+2|\Delta\nu|) \right\}^{1/2}
\cr
\nu_e &= q/2 \pm |\Delta\nu| \cr} \eqno{\hbox{(5-4)}}$$
Very close to the stopband edge, $|\nu_0-\nu_e|\ll|\Delta\nu|$, one finds
$$|\nu-q/2| = \left\{ 2|\Delta\nu| |\nu_0-\nu_e| \right\}^{1/2}
\eqno{\hbox{(5-5)}}$$
Thus, as $\nu_0$ approaches a stopband edge, $\nu$ approaches $q/2$,
and $d\nu/d\nu_0$ become infinite like $1/|\nu_0-\nu_e|^{1/2}$.

\section{The beta function near the edge of the stopband}

In this section, a result will be found for the beta function in the
stable region outside the stopband, but close to one of the edges of
the stopband.  It will be shown that close to edge of a stopband
$$[(\beta-\beta_0)/\beta_0]_{\rm max} = [2|\Delta\nu|/|\nu_0-\nu_e|]^{1/2}
\eqno{\hbox{(6-1)}}$$
$\nu_e$ is the edge of the stopband, $\nu_e=q/2\pm|\Delta\nu|$.  
$|\Delta\nu|$ is the half-width of the stopband.  $\nu_0$, $\beta_0$ are
the unperturbed tune and beta function.  Eq. (6-1) shows that when
$\nu_0$ approaches the edge of the stopband, $(\beta-\beta_0)/\beta_0$
becomes infinite like $1/|\nu_0-\nu_e|^{1/2}$.

The beta function $\beta$ can be found from the solution for the $\eta$
function which has the form given by Eq. (2-5).  It will be shown below
that $\beta$ is given by
$$\eqalign{
{\beta\over\beta_0} &= {\nu_0\over\nu} |\eta^2| \bigg\langle {1\over
|\eta|^2} \bigg\rangle \cr
\bigg\langle {1\over |\eta|^2} \bigg\rangle &= {1\over 2\pi} \int_0^{2\pi}
d\theta \ {1\over |\eta|^2} \cr} \eqno{\hbox{(6-2)}}$$
$\nu, \beta$ and $\nu_0, \beta_0$ are the perturbed and unperturbed
values of the tune and the beta function.  Thus $\beta$ can be determined 
from the $\eta$ function, if the perturbed tune is known, and in this case 
the tune was found in section 5.

To derive Eq. (6-2) one notes that
$$x=\beta_0^{1/2} \eta , \eqno{\hbox{(6-3)}}$$
and thus
$$\beta = D\ \beta_0 \ |\eta|^2$$
where $D$ is some constant.  The constant $D$ can be determined using
the relationship
$${\nu\over\nu_0} = {1\over 2\pi} \int_0^{2\pi} d\theta \ {\beta_0\over
\beta} \eqno{\hbox{(6-4)}}$$
which when written in this form, will hold also for small accelerators$^1$ 
when the perturbation does not change the closed orbit (see section 7).
  This gives
$$\eqalign{
D &= {\nu_0\over\nu} \Big\langle 1/|\eta|^2 \Big\rangle \cr
(\beta/\beta_0) &= (\nu_0/\nu) |\eta|^2 \Big\langle 1/|\eta|^2\Big\rangle
\cr} \eqno{\hbox{(6-5)}}$$
One now proceeds to find the $\eta$ function in the stable region near
an edge of the stopband.  To lowest order, $\eta$ is given by Eq. (3-7)
as
$$\eqalign{
\eta &= A_s \exp (i\nu_s\theta) + A_{\overline{s}} \exp(i\nu_{\overline{s}} 
\theta) \cr
\nu_{\overline{s}} &= \nu_s -q \cr} \eqno{\hbox{(6-6)}}$$
$A_{\overline{s}}$ is given by Eq. (3-9) which can be written as
$$\eqalign{
A_{\overline{s}} &= {|\Delta\nu|\exp(-i\delta_i)\over|\nu_s-q|-\nu_0}
A_{s} \cr
\delta_1 &= \hbox{ph} (\Delta\nu) \cr} \eqno{\hbox{(6-7)}}$$
Assuming $\nu_0$ is below the edge of the stopband one puts
$\nu_s=q/2-\delta$ and $\nu_0=\nu_e-\epsilon$, $\nu_e=q/2-|\Delta\nu|$.
Then
$$A_{\overline{s}} = {|\Delta\nu|\exp(-i\delta_1)\over \delta+\epsilon
+ |\Delta\nu|} A_s \eqno{\hbox{(6-8)}}$$
The $\eta$ functions is then given to lowest order by
$$\eqalign{
\eta &= A_s \exp(i\nu_s\theta) \left[ 1+{|\Delta\nu|\over |\Delta\nu|+
\epsilon+\delta} \exp [i(-q\theta-\delta_1)] \right] \cr
\epsilon &= |\nu_0-\nu_e| , \qquad \nu_e = q/2\pm |\Delta\nu| \cr
\delta &= |\nu-q/2| = \Big\{ |\nu_0-\nu_e| (|\nu_0-\nu_e|+2|\Delta\nu|)
\Big\}^{1/2} \cr} \eqno{\hbox{(6-9)}}$$
Eq. (6-9) has been written so that it holds both above and below the
stopband, and Eq. (5-4) was used to replace $\delta$.

One then finds
$$\eqalign{
|\eta|^2 &= |A_s|^2 (1+2 C \cos (q\theta+\delta_1) + C^2) \cr
C &= {|\Delta\nu|\over |\Delta\nu| + |\nu_0-\nu_e| + |\nu-q/2|} \cr}
\eqno{\hbox{(6-10)}}$$
$$\eqalign{
\Big\langle{1\over|\eta|^2}\Big\rangle &= {1\over |A_s|^2} {1\over 2\pi}
\int_0^{2\pi} d\theta {1\over 1+2C \cos(q\theta+\delta_1)+C^2} \cr
&= {1\over |A_s|^2} {1\over 1-C^2} \cr} \eqno{\hbox{(6-11)}}$$
One can now use Eq. (6-2) to find $\beta$
$${\beta\over\beta_0} = {\nu_0\over\nu} (1+C^2+2C \cos(q\theta+\delta_1))
{1\over 1-C^2} \eqno{\hbox{(6-12)}}$$
One may note that when $\nu_0$ is at the edge of a stopband,
$\nu_e-\nu_0=0$, $\nu-q/2=0$ and $C=1$.  Since one is interested here
in the rather large effects due to the $1/(1-C^2)$ factor near the edge
of the stopband, one can put $\nu_0/\nu=1$ with only a small error.
One finds for $\beta$
$$\eqalign{
(\beta-\beta_0)/\beta_0 &= {2C(C+\cos(q\theta+\delta_1))\over 1-C^2} \cr
((\beta-\beta_0)/\beta_0)_{\rm max} &= {2C\over 1-C} \cr
((\beta-\beta_0)/\beta_0)_{\rm max} &= {2|\Delta\nu|\over |\nu_0-\nu_e|
+ |\nu-q/2|} \cr} \eqno{\hbox{(6-13)}}$$
using the result $|\nu-q/2|=\{(\nu_0-\nu_e)(|\nu_0-\nu_e|+2|\Delta\nu|)
\}^{1/2}$ one gets
$$((\beta-\beta_0)/\beta_0)_{\rm max} = {2|\Delta\nu|\over |\nu_0-\nu_e| +
\{|\nu_0-\nu_e| (|\nu_0-\nu_e|+2|\Delta\nu|)\}^{1/2}} \eqno{\hbox{(6-14)}}$$
Very close to the stopband, $|\nu_0-\nu_e\ll|\Delta\nu|$, one finds
$$((\beta-\beta_0)/\beta_0)_{\rm max} = \left\{ {2|\Delta\nu|\over
|\nu_0-\nu_e|}\right\}^{1/2} \eqno{\hbox{(6-15)}}$$
Eq. (6-15) shows that as $\nu_0$ approaches the edge of a stopband
$((\beta-\beta_0)/\beta_0)_{\rm max}$ becomes infinite like
$\{|\nu_0-\nu_e|\}^{-1/2}$.

Table 1 below shows how the distortion in the beta function depends
on how far $\nu_0$ is from the edge of stopband. The table shows
that to keep $((\beta-\beta_0)/\beta_0)_{\rm max}$ less than 10\%,
then $\nu_0$ has to be about $10\Delta\nu$ away from the stopband
edge.

Table 1:  $((\beta-\beta_0)/\beta_0)_{\rm max}$ versus $|\nu_0-\nu_e|/
|\Delta\nu|$ as computed from Eq. (6-14).

\medskip

\ruledtable
$\vert\nu_0-\nu_e\vert/\vert\Delta\nu\vert$ & .5 & 1 & 2 & 4 & 6 & 8 & 10 \crnorule
$((\beta-\beta_0)/\beta_0)_{\rm max}$ & 1.24 & .73 & .41 & .22 & .15 &
.12 & .095 \endruledtable

\section{Comments on the small accelerator results}

All the final results found in this paper will also hold for small
accelerators where the exact equations of motions have to be used.
The exact linear equations have the form$^1$
$$\eqalign{
{dx\over ds} &= A_{11} x + A_{12} q_x \cr
{dq_x\over ds} &= A_{21} x + A_{22} q_x \cr
q_x &= p_x/p \cr} \eqno{\hbox{(7-1)}}$$
In the large accelerator approximation, it is assumed that $A_{11} =
A_{22} = 0$, $A_{12} = 1$, and $q_x\simeq dx/ds$.  The coefficients
$A_{ij}$ are given in reference 1.  In particular, $A_{12}$ is
given by
$$A_{12} = {1+x/\rho\over (1-q_x^2)^{3/2}}. \eqno{\hbox{(7-2)}}$$
where the right hand side is evaluated on the closed orbit.

Although for small accelerators, the differential equations for $x$ and
$\beta$ are different from those usually used for large accelerators,
it has been found$^1$ that the linearized differential equation for
$\eta = x/\beta^{1/2}$ is not much different.  One finds
$$\eqalign{
{d^2\over d\theta^2} \eta &+ \nu_0^2 \eta = f \cr
f &= {\nu_0^2\beta_0^2\over A_{12}} {\Delta B_y\over B\rho} \cr
B\rho &= pc/e \cr} \eqno{\hbox{(7-3)}}$$
with the difference that $\theta$ is now defined by$^1$
$$d\theta = A_{12} {ds\over\nu_0\beta_0} \eqno{\hbox{(7-4)}}$$
where $\nu_0,\beta_0$ are the unperturbed tune and beta function.
The $A_{12}$ is evaluated on the unperturbed closed orbit.

The relationshp between $\nu$ and $\beta$ is somewhat different
for small accelerators, and is given by$^1$
$$\nu = {1\over 2\pi} \int_0^C {A_{12}\over\beta} ds \eqno{\hbox{(7-5)}}$$
where $C$ is the circumference of the accelerator.  Using Eq. (7-4) this
can be written as
$${\nu\over\nu_0} = {1\over 2\pi} \int_0^{2\pi} {A_{12}\over \overline{A}_{12}}
{\beta_0\over\beta} d\theta \eqno{\hbox{(7-6)}}$$
where $\overline{A}_{12}$ is the $A_{12}$ coefficient evaluated on the
unperturbed closed orbit.  If the perturbation being considered does not
change the closed orbit, then $A_{12} = \overline{A}_{12}$ and one has
$${\nu\over\nu_0} = {1\over 2\pi} \int_0^{2\pi} {\beta_0\over\beta} 
d\theta \eqno{\hbox{(7-7)}}$$
Eq. (7-7) now holds for both large and small accelerators, provided
the perturbation does not change the closed orbit.

The results found in this paper are based on Eq. (7-3) for $\eta$ and
Eq. (7-7) that relates $\nu$ and $\beta$.  Although, the perturbation
term in Eq. (7-3) now has the extra factor $1/A_{12}$ for small
accelerators, this factor of $1/A_{12}$ disappears in the final
result when one goes from the variable $\theta$ to the variable $s$
according to Eq. (7-4).  Thus the result for the stopband width,
$\Delta\nu$, as given by Eq. (3-11b) is valid for both large and
small accelerators.  Keeping the above equations in mind, one can
go through the derivations on the previous pages, and show that the
final results are valid for both large and small accelerators.

\section{Comments on the results}

Others have worked on this subject and there is an  overlap between
the contents of this paper and their work. These previous papers (2 to 6 )  
give results for the stopband width and for the growth rate inside the
stopband.

The results in this paper include the following.  The result given
for the tune $\nu$ near the edge of the stopband, $\nu_e$, $|\nu-q/2| =
[2|\Delta\nu||\nu_e-\nu_0|]^{1/2}$.  The result given for the beta
function near the edge of the stopband, $[(\beta-\beta_0)/\beta_0]_{\rm max}
= [2|\Delta\nu|/|\nu_e-\nu_0|]^{1/2}$.
The proof given showing that the real part of $\nu$
is constant over the stopband at $q/2$ does not depend on perturbation
theory, and the result follows from the symplectic properties.  The result 
that all the results found in this paper will also hold for a small accelerator
where the large accelerator approximation is not used.  The result given
for the solutions of the equations of motion when $\nu_0$ is inside the
stopband, and the proof that the eigenfunctions and eigenvalues of the
transfer matrix are real inside the stopband.

\nosechead{\tbf References}

\item{1.} G. Parzen, Linear orbit parameters for the exact equations
of motion, BNL Report, BNL-60090 (1994).
\item{2.} P.A. Sturrock, Static and dynamic electron optics, Cambridge
Univ. Press, London (1955).
\item{3.} E.D. Courant and H.S. Snyder, Theory of the alternating
gradient synchrotron, Annals. of Physics, Vol. 3, p. 1 (1958).
\item{4.} H. Bruck, Circular particle accelerators, (1965), English
translation Los Alamos Report, LA-TR-72-10 Rev.
\item{5.} A.A. Kolomensky and A.N. Lebedev, Theory of cyclic accelerators,
North Holland Publishing Co., Amsterdam (1966).
\item{6.} H. Wiedemann, Particle accelerator physics, Springer-Verlag,
Berlin (1993).

\bye